\documentclass[aps,pra,reprint,superscriptaddress,showpacs]{revtex4-1}
\usepackage{graphicx,amsmath,amssymb,ulem}
\usepackage{verbatim}
\usepackage{ragged2e}
\usepackage{color}
\usepackage{color, colortbl}
\usepackage{physics}
\usepackage{csquotes}
\usepackage[section]{placeins}
\usepackage{ulem}

\makeatletter
\renewcommand{\fnum@figure}{Fig. \thefigure}
\makeatother

\makeatletter
\renewcommand{\fnum@table}{Table \thetable}
\makeatother

\renewcommand{\thetable}{\arabic{table}}   

\newcommand{\RomanNumeralCaps}[1]
    {\MakeUppercase{\romannumeral #1}}

\definecolor{LightCyan}{rgb}{0.88,1,1}

\definecolor{orange}{RGB}{255,127,0}

\begin{document}
\title{Generating two-mode squeezing with multimode measurement-induced nonlinearity}
\author{M.~Riabinin}
\affiliation{Department of Physics, University of Paderborn, Warburger Stra\ss{}e 100, D-33098 Paderborn, Germany}
\author{P.~R.~Sharapova}
\affiliation{Department of Physics, University of Paderborn, Warburger Stra\ss{}e 100, D-33098 Paderborn, Germany}
\author{T.~J.~Bartley}
\affiliation{Department of Physics, University of Paderborn, Warburger Stra\ss{}e 100, D-33098 Paderborn, Germany}
\author{T.~Meier}
\affiliation{Department of Physics, University of Paderborn, Warburger Stra\ss{}e 100, D-33098 Paderborn, Germany}


\begin{abstract}
Measurement-induced nonclassical effects in a two-mode interferometer are investigated theoretically using numerical simulations and analytical results. We demonstrate that for certain parameters measurements within the interferometer lead to the occurrence of two-mode squeezing. The results strongly depend on the detection probability, the phase inside the interferometer, and the choice of the input states. The appropriate parameters for maximized squeezing are obtained. We analyze the influence of losses and confirm that the predicted effects are within reach of current experimental techniques.
\end{abstract}
\maketitle

\section{INTRODUCTION}In continuous-variable quantum optics, quadrature squeezing and non-Gaussian entanglement are two important properties of nonclassical light which are required for the implementation of many quantum computation and  communication protocols~\cite{CV-QIP}. {The most common method used} to generate quadrature squeezed light is to {exploit} a nonlinear interaction in the medium~\cite{Klyshko}, {for example} by parametric down-conversion (PDC) or four-wave mixing (FWM). In the low-gain regime, the amount of squeezing in such processes is proportional to the intensity of the pump fields \cite{Andersen}. 
{In some cases, however, it may be desirable to generate squeezing and entanglement without a strong pump.} 
An alternative 
is {to use} measurement-induced nonlinearities (MINL), whereby nonlinear effects can be acquired by applying detection~\cite{Mes-ind-nonlin, Kalman}. 

Early experimental work from Lvovsky and Mlynek showed that combining measurement-induced nonlinearity with single-photon ancilla states, through a process they termed ``quantum catalysis,'' a number of nonclassical properties can be induced~\cite{Lvovsky_Mlynek_quant_catal}. Since then, combining single-photon ancilla states with single photon measurement has been used for further exotic state generation and manipulation~\cite{Usuga, Marek1, Muller, Marek2, Miyata, Yukawa, Marek3, Ourjoumtsev, Ferreyrol, Xiang, Sanaka, Resch, Bartley}. 
In the context of quadrature squeezing, 
it was shown that, depending on the interaction parameters, the state in the single output mode may be squeezed~\cite{Bartley}. Although the amount of squeezing is limited to 1.25 dB, the appearance of single-mode quadrature squeezing from conditional interference of a single photon and a weak coherent state is not immediately intuitive.
The question therefore arises whether, when expanding to more modes, two-mode squeezing~\cite{Han, Pogorzalek, Diniz, Lawrie, Rojas, Larsen, Magana-Loaiza} can be induced using a similar scheme and more generally, whether other classes of multimode entangled states can be generated. Such studies are interesting in the context of identifying the resource requirements for generating multimode non-Gaussian states, which are required for a wide range of continuous-variable quantum information protocols~\cite{CV-QIP}.

In this work, we present a theoretical investigation of measurement-induced nonlinearity in a four-mode system. We consider a two-mode interferometer in which the single-photon measurements occur within the interferometer itself. 
In the resulting two output channels, we analyze the acquired nonclassical effects  conditional on certain detection events. It is shown that the implemented detection modifies the photon statistics and leads to two-mode squeezing (TMS) in the system.

This paper is organized as follows: In section \RomanNumeralCaps{2} we present our theoretical description of the scheme. In section \RomanNumeralCaps{3} we present and discuss the analytical results and numerical simulations which demonstrate squeezing for optimized parameters in the case of photon-number-resolved detection. In section \RomanNumeralCaps{4} we consider the case of click detection. In section \RomanNumeralCaps{5} we visualize the generated states with their Wigner functions. In section \RomanNumeralCaps{6} we consider the influence of losses. We close with a brief summary in section \RomanNumeralCaps{7}. Additional analytical results are provided in the Appendix~\ref{appa}.

\section{THEORETICAL MODEL}

\begin{figure}[htb]
\begin{center}
\includegraphics[width=0.49\textwidth]{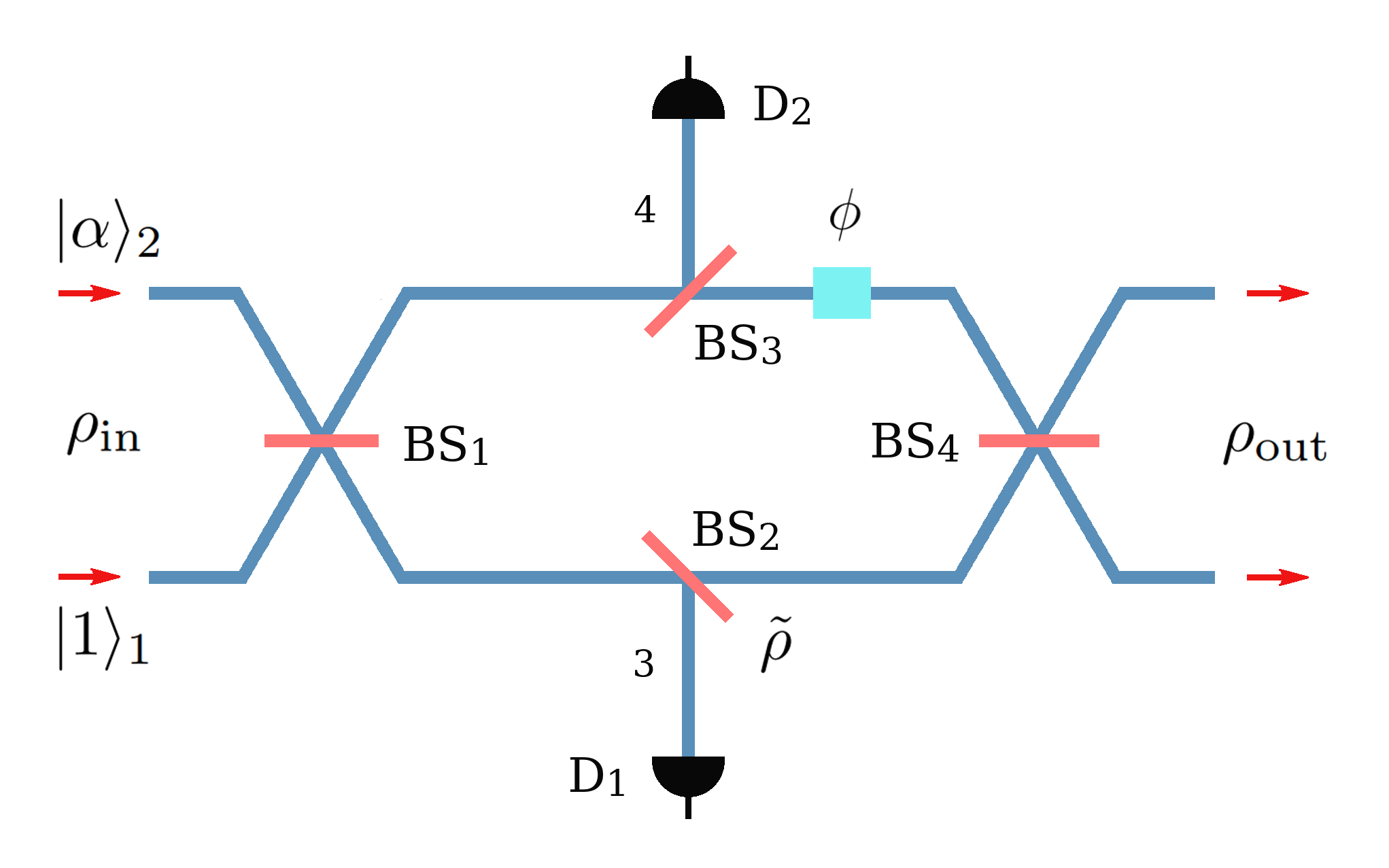}
\end{center}
\caption{A schematic representation of the considered interferometer comprising four beam splitters $\textrm{BS}_{1}$-$\textrm{BS}_{4}$ (red), a phase shifter (cyan), and two detectors $\textrm{D}_{1}$ and $\textrm{D}_{2}$. $\textrm{BS}_{2}$ and $\textrm{BS}_{3}$ model the partial out-coupling of light from the interferometer to the detectors.} \label{fig:full_setup}
\end{figure}

The scheme we consider is shown in Fig.~\ref{fig:full_setup}. We consider a Mach-Zehnder interferometer, into which various quantum states can be injected. We consider the specific case of input states which do not exhibit (single-mode) quadrature squeezing, namely a single photon state $|1\rangle_{1} = \hat{a}_{1}^{\dag} |0\rangle$ in channel 1 and a coherent state of mean photon number $|\alpha|^2$, $|\alpha\rangle_{2} = \exp(-\frac{1}{2}|\alpha|^2)  \sum_{n=0} \frac{\alpha^{n}}{n!}(\hat{a}_{2}^{\dag})^{n} |0\rangle$ in channel 2 respectively, as shown in Fig.~\ref{fig:full_setup}. 
The quantum state of light injected in the two channels is thus the tensor product $|\psi\rangle = |1\rangle_{1}  \otimes |\alpha\rangle_{2}$.

The state following interference of the single photon and the coherent state at the first beam splitter is entangled, and exhibits interesting photon statistics \cite{Biagi, DiscQStates}. However, the output state does not exhibit two-mode quadrature squeezing, for any beamsplitter parameter chosen. Two-mode squeezing can nevertheless be induced by certain outcomes of measurements made within the interferometer.

The whole interferometer acts on the input light as a series of transformations. First, the light passes through $\textnormal{BS}_{1}$.
Afterwards it is split up into the four channels at the beam splitters  $\textnormal{BS}_{2}$ and $\textnormal{BS}_{3}$.  Detection is possible in the channels 3 and 4, governed by the detection operator $\hat{\mathrm{D}}$. The explicit form of $\hat{\mathrm{D}}$ depends on the type of detector used, as described in the following section. 
After detection, a phase shift $\hat{\textnormal{P}}_{2}(\phi)$ is implemented in the upper channel before the final beam splitter $\textnormal{BS}_{4}$.
Altogether, the resulting transformations defining the relation between the input and output density matrices can be written as:

\begin{equation} \label{eq:full_transformation}
\begin{split}
& \tilde{\rho} = 
\textnormal{Tr}_{3,4}[
\hat{\textnormal{\bf B}}\left(t_{2}, t_{3}\right)
\hat{\textnormal{B}}\left(t_{1}\right)
\rho_{\mathrm{in}}
\hat{\textnormal{B}}^\dag\left(t_{1}\right)
\hat{\textnormal{\bf B}}^\dag\left(t_{2}, t_{3}\right)
\hat{\textnormal{D}}] \\
& \rho_{\mathrm{out}} = 
\hat{\textnormal{B}}\left(t_{4}\right)
\hat{\textnormal{P}}_{2}\left(\phi\right)
\tilde{\rho}
\hat{\textnormal{P}}_{2}^\dag\left(\phi\right)
\hat{\textnormal{B}}^\dag\left(t_{4}\right)
\end{split}
\end{equation}

In this formalism, the evolution of the input states through the interferometer in Fig.~\ref{fig:full_setup} is characterized by a set of (lossless) beamsplitters BS$_i$ of transmission $T_i = t_i^2 = \cos^2\left(\theta_{i}\right)$ and reflection  $R_i = r_i^2 = 1 - T_i = \sin^2\left(\theta_{i}\right)$ coefficients. In general, the BS operator $\hat{\textnormal{B}}\left(t_i\right) = \exp\left[ i \theta_i \left( \hat{a}_1 \hat{a}_2^{\dag} +  \hat{a}_1^{\dag} \hat{a}_2\right) \right]$ with $t_i=\cos\left(\theta_i\right)$
defines a linear transformation of the creation (and annihilation) operators between the input $\textbf{\textit{a}}^{\dag} = \left[\hat{a}_{1}^{\dag}, \hat{a}_{2}^{\dag}\right]^{T}$ and the output $\textbf{\textit{b}}^{\dag} = \left[\hat{b}_{1}^{\dag}, \hat{b}_{2}^{\dag}\right]^{T}$ modes~\cite{Loudon}:
\begin{equation} \label{eq:BS transform.}
\begin{split}
\hat{\textnormal{B}}\left(t\right) \hat{a}_{1}^\dagger \hat{\textnormal{B}}^\dagger\left(t\right) &= \hat{b}_{1}^\dagger\cos \left(\theta\right) - i \hat{b}_{2}^\dagger  \sin \left(\theta\right)  = t \hat{b}_{1}^\dagger - i r \hat{b}_{2}^\dagger,\\
\hat{\textnormal{B}}\left(t\right) \hat{a}_{2}^\dagger \hat{\textnormal{B}}^\dagger\left(t\right) &= \hat{b}_{2}^\dagger\cos\left(\theta\right) - i \hat{b}_{1}^\dagger   \sin \left(\theta\right)= t \hat{b}_{2}^\dagger - i r \hat{b}_{1}^\dagger.
\end{split}
\end{equation}
The density matrix of the quantum state can be written as a function of creation and annihilation operators $ \rho = \rho\left(\textbf{\textit{a}}, \textbf{\textit{a}}^{\dag}\right)$.
The transformation of the density matrix at each BS can be obtained by using the input/output relations, Eq.~(\ref{eq:BS transform.}), for each operator.
The output density matrix $\rho_{a}$ is obtained from input $\rho_{b}$ by applying the BS transformation operator $\hat{\textnormal{B}}\left(t\right)$, {i.e.} $\rho_{a} = \hat{\textnormal{B}}\left(t\right) \rho_{b} \hat{\textnormal{B}}^\dagger\left(t\right)$.
The action of $\textnormal{BS}_{2}$ and $\textnormal{BS}_{3}$ is considered together by the operator $\hat{\textnormal{\bf B}}(t_{2}, t_{3}) = \hat{\textnormal{B}}(t_{2}) \hat{\textnormal{B}}(t_{3})$ which describes a transformation of the operators $\hat{a}_{1}^{\dag}$ and $\hat{a}_{2}^{\dag}$ into output modes $\hat{b}^{\dag}$ in four channels:
\begin{equation} \label{eq:BS_23.}
\begin{split}
\hat{\textnormal{B}}\left(t_{2}\right)\hat{a}_{2}^\dagger \hat{\textnormal{B}}^\dagger\left(t_2\right) & = t_2 \hat{b}_{2}^\dagger - i r_2 \hat{b}_{3}^\dagger, \\
\hat{\textnormal{B}}\left(t_{3}\right)\hat{a}_{1}^\dagger \hat{\textnormal{B}}^\dagger\left(t_3\right) & = t_3 \hat{b}_{1}^\dagger - i r_3 \hat{b}_{4}^\dagger, 
\end{split}
\end{equation}
where $\hat {\textnormal{B}}\left(t_{2}\right)=\exp\left[ i \theta_{2} \left( \hat{a}_2 \hat{a}_3^{\dag} +  \hat{a}_2^{\dag} \hat{a}_3\right) \right]$, $\hat {\textnormal{B}}\left(t_{3}\right) = \exp\left[ i \theta_{3} \left( \hat{a}_1 \hat{a}_4^{\dag} +  \hat{a}_1^{\dag} \hat{a}_4\right) \right]$, and, as before, $t_i = \cos{\theta_i}$.



\subsection{Detectors}
In this work we consider two types of detectors: click detectors which measure the absence or presence of photons but provide no information about the photon number and photon-number-resolving (PNR) detectors~\cite{Det_quantum_light}.

PNR detection in one channel can be described by the projection of the state on the chosen Fock state with $n$ photons in the $i$-th channel $|n\rangle_{i}$: $|\psi_{\mathrm{a}}\rangle=  |n\rangle\langle n|_{i} | \psi_{\mathrm{b}}\rangle$ where $|\psi_{\mathrm{b}}\rangle$ is the state before detection. 
The probability of such an event is $P_{\mathrm{det}} = \langle\psi_{\mathrm{a}}|\psi_{\mathrm{a}}\rangle$, where $|\psi_{\mathrm{a}}\rangle$ is the unnormalized state after projection. 
For simplicity, we consider only single-photon PNR detection, i.e., projection onto the single photon state $|1\rangle $ or the vacuum state $|0\rangle $, in this work.
Therefore, for the two detectors in channels 3 and 4, see Fig.~\ref{fig:full_setup}, we consider four different outcomes: (i) both detectors measure one photon, (ii) only the detector in channel 4 registers a photon, (iii) only the  detector in channel 3 registers a photon, and (iv) both detectors measure vacuum:
\begin{equation} \label{eq:det_spd}
\begin{aligned}
&|\psi_{\mathrm{a}}^{(3\&4)}\rangle&=& \ |1\rangle\langle 1|_{3} \otimes |1\rangle\langle 1|_{4}  |\psi_{\mathrm{b}}\rangle \\ 
&|\psi_{\mathrm{a}}^{(4)}\rangle&=& \ |0\rangle \langle 0|_{3} \otimes |1\rangle \langle 1|_{4} |\psi_{\mathrm{b}}\rangle \\ 
&|\psi_{\mathrm{a}}^{(3)}\rangle&=& \ |1\rangle \langle 1|_{3} \otimes |0\rangle \langle 0|_{4} |\psi_{\mathrm{b}}\rangle \\ 
&|\psi_{\mathrm{a}}^{(\mathrm{none})}\rangle&=& \ |0\rangle \langle 0|_{3} \otimes |0\rangle \langle 0|_{4} |\psi_{\mathrm{b}}\rangle, \\ 
\end{aligned}
\end{equation}
where $|\psi_{\mathrm{b}}\rangle$ is the state in the four channels after $\textnormal{BS}_{2}$ and $\textnormal{BS}_{3}$ but before detection. 

By contrast, click detectors do not resolve the number of photons and must therefore take into account all possible photon-number contributions. The action of click detectors can be described in terms of the positive operator valued measure (POVM) operators $ \hat{\Pi}^{(-)} = |0\rangle  \langle 0|$ and $\hat{\Pi}^{(+)} = \hat{\mathrm{I}} - \hat{\Pi}^{(-)} = \sum_{n=1}^{\infty}|n\rangle  \langle n| $ which describe the absence and presence of a click, respectively. The two detectors are again described by four possible projection operators:
\begin{equation} \label{eq:det_povm_oper2}
\begin{aligned}
&\hat{\Pi}_{3\textrm{\&}4}&=& \ \hat{\Pi}_{3}^{(+)} \otimes  \hat{\Pi}_{4}^{(+)}  \\
&\hat{\Pi}_{4}&=& \ \hat{\Pi}_{3}^{(-)} \otimes  \hat{\Pi}_{4}^{(+)} \\ 
&\hat{\Pi}_{3}&=& \ \hat{\Pi}_{3}^{(+)} \otimes  \hat{\Pi}_{4}^{(-)} \\
&\hat{\Pi}_{\mathrm{none}}&=& \ \hat{\Pi}_{3}^{(-)} \otimes  \hat{\Pi}_{4}^{(-)} .
\end{aligned}
\end{equation}

To obtain a density matrix after detection $\rho^\prime$, in the click detection case we apply the POVM operators to the density matrix before detection $\rho$ and take the partial trace over the detecting channels $ \rho^\prime =  \mathrm{Tr}_{3,4}(\rho  \hat{\Pi}_{\mathrm{event}})$. The detection probability is given by $ P_{\mathrm{det}}^{(\mathrm{event})} = \mathrm{Tr}(\rho^\prime  \hat{\Pi}_{\mathrm{event}})$, where $\hat{\Pi}_{\mathrm{event}}$ is the POVM for a particular measurement event. 

For both PNR and click detection, each measurement outcome leads to an unnormalized density matrix.
Therefore, we define a new normalized detection operator $\hat{\textnormal{D}}=\hat{\Pi}_{\mathrm{event}}/ P_{\mathrm{det}}^{(\mathrm{event})}$.

\subsection{Generating two-mode squeezing}
Single-mode squeezing is defined as the reduction of the quadrature variance below the shot noise level $\Delta^{2} X_{i} < \frac{1}{4}$
~\cite{ScullyZubairy, LoudonKnight}, with generalized quadratures defined by
$X_{1}^a (\xi) = \frac{1}{2}(e^{-i\xi}\hat{a} + e^{i\xi}\hat{a}^\dag) $ and $ X_{2}^a (\xi) = \frac{1}{2i}(e^{-i\xi}\hat{a} - e^{i\xi}\hat{a}^\dag) $ where $\xi$ is a quadrature phase
and the variance is defined as $ \Delta^{2} X = \langle X^2 \rangle - \langle X \rangle^2$. TMS between modes $a$ and $b$ is connected with the mutual variance of quadratures and is described by the joint quadrature operators~\cite{Schnabel}:

\begin{equation}  \label{eq:quadr_definition1}
\begin{aligned}
& C_{1} = \frac{1}{\sqrt{2}}(X_{1}^a + X_{1}^b) = \frac{1}{\sqrt{8}}(e^{-i\xi}(a + b) + e^{i\xi}(a^\dag + b^\dag)) \\
& C_{2} = \frac{1}{\sqrt{2}}(X_{2}^a + X_{2}^b) = \frac{1}{i\sqrt{8}}(e^{-i\xi}(a + b) - e^{i\xi}(a^\dag + b^\dag)) \\
\end{aligned}
\end{equation}

Similarly to the single-mode case, two-mode light is squeezed if one of the variances in Eq.~(\ref{eq:quadr_definition1}) is lower than the shot noise level: $ \Delta^{2} C_{i} < \Delta^{2} C_{i}^{(0)} = \frac{1}{4} $. The condition $ \Delta^{2} C_{1} = \frac{1}{2}\Delta^{2} X_{1}^{a} +  \frac{1}{2}\Delta^{2} X_{1}^{b} + \mathrm{Cov}[X_{1}^{a}, X_{1}^{b}] < \frac{1}{4}$ can be satisfied either if the two modes are uncorrelated and, simultaneously, one or both of them are individually squeezed, or when nonclassical correlations between the modes (entanglement) exist. Two-mode squeezing is defined as a reduction of the variance in comparison to the shot noise level $S_{i} = 10 \textnormal{log}_{10}(\Delta^{2} C_{i}/\Delta^{2} C_{i}^{(0)})$.

\subsection{Role of detection}
To demonstrate the significance of the detectors for the generation of two-mode squeezing in the circuit depicted in Fig.~\ref{fig:full_setup}, we first investigate a simplified setup without detection, i.e.,  the beam splitters $\mathrm{BS}_{2}$ and $\mathrm{BS}_{3}$ transmit the light with $T_{2}=T_{3}=1$. Considering the input state $|1 \rangle_1   \otimes |\alpha \rangle_{2} $, the variances can be calculated analytically: $\Delta^{2} C_{1} = \Delta^{2} C_{2} = \frac{1}{2}  + \sin{\phi} (\frac{1}{2}(t_{1}r_{1} + t_{4}r_{4}) - t_{1}^2t_{4}r_{4} - t_{4}^2t_{1}r_{1})$. This result depends neither on the mean number of photons of the coherent state $\alpha$ nor on the quadrature phase $\xi$ and has a minimum of $\Delta^{2} C_{i} =\frac{1}{4}$ (0 dB). This means that in the case of only linear elements and non-squeezed input states, the output light is not squeezed as well.

\subsubsection{Numerical optimization routine}\label{sec:optimisation}
Adding detectors to the scheme may generate two-mode squeezing.
To demonstrate this in general, we
 perform a numerical optimization to find the minimum of the variance $\Delta^{2} C_{i}$ in order to maximize squeezing. We use the following algorithm: chose some detection event $d=D_{j}$, constrain the probability to be higher than some minimum value $P_{\mathrm{crit}}$, fix phases $\phi$ and $\xi$, and then minimize the variance over all BS parameters:
\begin{equation}  \label{eq:epr_minimization}
\begin{aligned}
& \text{minimize:}
& & \Delta^{2} C_{i}(\textbf{T}, \phi, \xi, d) \\
& \text{subject to:}
& &  \phi=\phi_{0}, \\  &&& \xi = \xi_{0}, \\ &&& d = D_{j} , \\ &&& P_{\mathrm{det}} \geq P_{\mathrm{crit}} 
\end{aligned}
\end{equation}
The constraint on the probability is implemented to avoid cases where squeezing may be generated with vanishing detection probability. In principle, arbitrary values for $P_\textrm{crit}$ may be chosen; this will be determined by the parameters of an experiment.

Although the quantities $\Delta^{2} C_{1}$ and $\Delta^{2} C_{2}$  exhibit smooth continuous behavior over all parameters, it is still numerically difficult to find a global minimum of these four-variable functions. The straightforward approach with evaluating variances over a multidimensional grid and choosing their minimal values is computationally expensive. One way to improve the situation is to use a gradient descent-based algorithms. In this work we apply the gradient-based algorithm "Adam"~\cite{Adam} with its TensorFlow library implementation~\cite{tensorflow}. To speed up the convergence of the algorithm, different starting points were chosen. 


\section{Photon-number-resolved detection}\label{sec:PNR}

For the case of PNR detection, analytical expressions for the output states and the detection probabilities can be obtained
for the cases of single- and both-channel detection which are given in Eqs.~(\ref{eq:state_det_pnr_single})-(\ref{eq:prob_pnr_det_both}). For the case of single-channel PNR detection the state is given by:
\begin{equation}  \label{eq:state_det_pnr_single}
\begin{aligned}
  \rho_{\textrm{out}} =& | \psi_{\mathrm{single}} \rangle \langle \psi_{\mathrm{single}}  | \\
| \psi_{\mathrm{single}} \rangle =& N_{\mathrm{single}}\left(\gamma_{0} + \gamma_{1}\hat{a}_{1}^{\dag} + \gamma_{2}\hat{a}_{2}^{\dag}\right)| \alpha_{1}, \alpha_{2} \rangle \\
 N_{\mathrm{single}} =& \tilde{P}_{\mathrm{single}}^{-\frac{1}{2}} \exp\left[-\frac{1}{2}\left(\left|\alpha_3\right|^2+\left|\alpha_4\right|^2\right)\right]~, 
\end{aligned}
\end{equation}
with coefficients
\begin{equation}\nonumber
\begin{aligned}
&  \gamma_{0} = i t_{1} r_{3}  \\
&  \gamma_{1} = \alpha_{\textrm{in}} r_{1} r_{3} \left ( r_{1} t_{2} t_{4} - t_{1} t_{3} r_{4} e^{i\phi}   \right  ) \\
&  \gamma_{2} = -i \alpha_{\textrm{in}} r_{1} r_{3} \left ( r_{1} t_{2} t_{4} + t_{1} t_{3} r_{4} e^{i\phi}   \right )  \\
&  \alpha_{1} = i \alpha_{\textrm{in}} \left(t_{1} t_{2} r_{4} + r_{1} t_{3} t_{4} e^{i\phi} \right ) \\
&  \alpha_{2} = \alpha_{\textrm{in}} \left(t_{1} t_{2} t_{4} - r_{1} t_{3} r_{4} e^{i\phi} \right  ) \\
&  \alpha_{3} = i \alpha_{\textrm{in}} t_{1} r_{2}  \\
&  \alpha_{4} = - \alpha_{\textrm{in}} r_{1} r_{3}~, \\
\end{aligned}
\end{equation}
where $|\alpha_{1}, \alpha_{2} \rangle = |\alpha_{1}\rangle_{1} \otimes |\alpha_{2} \rangle_{2}$ is the product of two coherent states and $ \alpha_{\textrm{in}}$ is the initial coherent state.

Following a detection event, the state after detection is not normalized. To normalize this state we use the detection probability; or the state given by Eq.~(\ref{eq:state_det_pnr_single}) this probability is given by:
\begin{equation}  \label{eq:prob_pnr_det_single}
\begin{aligned}
 \tilde{P}_{\mathrm{single}} =& \exp\left[-|\alpha_{\textrm{in}}|^2 \left(T_{1}R_{2} + R_{1}R_{3}\right)\right] \\
& \times R_{3} [|\alpha_{\textrm{in}}|^2 R_{1}^2 T_{2} + T_{1}(1 + |\alpha_{\textrm{in}}|^2 R_{1}(3T_{3} - 2T_{2}) \\
& + |\alpha_{\textrm{in}}|^4 R_{1}^2(T_{2} - T_{3})^2)] .
\end{aligned}
\end{equation}

For the case where both channels register a detection event the output state is given by:
\begin{equation}  \label{eq:state_det_pnr_both}
\begin{aligned}
  \rho_{\textrm{out}} = &| \psi_{\mathrm{both}} \rangle \langle \psi_{\mathrm{both}}  | \\
| \psi_{\mathrm{both}} \rangle =& N_{\mathrm{both}}\left(\tilde{\gamma}_{0} + \tilde{\gamma}_{1}\hat{a}_{1}^{\dag} + \tilde{\gamma}_{2}\hat{a}_{2}^{\dag}\right)| \alpha_{1}, \alpha_{2} \rangle
\end{aligned}
\end{equation}
where
\begin{equation}\nonumber
\begin{aligned}
& N_{\mathrm{both}} = \tilde{P}_{\mathrm{both}}^{-\frac{1}{2}} \exp\left[-\frac{1}{2}\left(\left|\alpha_3\right|^2+\left|\alpha_4\right|^2\right)\right] \\
&  \tilde{\gamma}_{0} = \alpha_{\textrm{in}} r_{2} r_{3} \left(2r_{1}^2 - 1 \right)  \\
&  \tilde{\gamma}_{1} = \gamma_{1} \alpha_{3} \\
&  \tilde{\gamma}_{2} = \gamma_{2} \alpha_{3}\end{aligned}
\end{equation}
and the probability of realizing the state in Eq.~(\ref{eq:state_det_pnr_both}) is:
\begin{equation}  \label{eq:prob_pnr_det_both}
\begin{aligned}
\tilde{P}_{\mathrm{both}} = &\exp\left[-\left|\alpha_{\textrm{in}}\right|^2 \left(T_{1}R_{2} + R_{1}R_{3}\right)\right] \\
& \times \left|\alpha_{\textrm{in}}\right|^2 R_{2} R_{3}\bigg\{T_{1}^2 + R_{1}^2\big[1 + \left|\alpha_{\textrm{in}}\right|^2 T_{1}\left(3T_{2} - 2T_{3}\right) \\
& + \left|\alpha_{\textrm{in}}\right|^4 T_{1}^2\left(T_{2} - T_{3}\right)^2\big] \\
& + R_{1}\left[-2 T_{1} + \left|\alpha_{\textrm{in}}\right|^2T_{1}^2\left(-2T_{2} + 3T_{3}\right)\right]\bigg\}~.
\end{aligned}
\end{equation}

The formulas for the output density matrices for the  single-detector-click and the both-detectors-click for PNR detection cases share a similar form, but with different sets of coefficients $\gamma_{i}$ and $\tilde{\gamma}_{i}$, see Eqs.~(\ref{eq:state_det_pnr_single}) and (\ref{eq:state_det_pnr_both}). However, the probabilities for detecting one and two photons in the system are different, see Eqs.~(\ref{eq:prob_pnr_det_single}) and (\ref{eq:prob_pnr_det_both}).

\subsection{Special case: Neglecting BS$_4$}
To analyze the analytical expressions above, we start by considering a particular case of a simplified interferometer with fixed parameters $T_{1} = \frac{1}{2}, \ T_{2}=T_{3} \equiv T \ \textrm{and} \ T_{4} = 1$. Using Eqs.~(\ref{eq:quadr_variances}) and (\ref{eq:oper_averages}) from the Appendix, the analytical results for variances for the single PNR detection case take the form

\begin{equation}  \label{eq:quadr_definition}
\begin{aligned}
\Delta^2 C_{1} = &\frac{1}{\left(1 + x\right)^{2}} \bigg\{\frac{1}{4} + \frac{x}{2} + \frac{x^2}{2} + \frac{x}{8} \big[-\cos\left(2\xi\right) +  \\
& + \cos\left(2\xi - 2\phi\right) + 2\sin\left(2\xi - \phi\right) + 2 x \sin\left(\phi\right)\big]\bigg\} \\
\Delta^2 C_{2} = &\frac{1}{\left(1 + x\right)^{2}} \bigg\{\frac{1}{4} + \frac{x}{2} + \frac{x^2}{2} + \frac{x}{8}\big[\cos\left(2\xi\right) -  \\
& - \cos\left(2\xi - 2\phi\right) - 2\sin\left(2\xi - \phi\right) + 2 x \sin\left(\phi\right)\big]\bigg\} \\
 x =& T \alpha^2, \\
\end{aligned}
\end{equation}

with the probability of detection given by

\begin{equation}  \label{eq:simlf_det_prob}
\begin{aligned}
&  \tilde{P}_{\textnormal{single}} = \frac{1-T}{2} \exp\left[-|\alpha|^2 \left(1-T\right)\right] \left(1 + |\alpha|^2 \frac{3T}{2} \right)~.  \\
\end{aligned}
\end{equation}

\begin{figure}
\begin{center}
\includegraphics[width=0.46\textwidth]{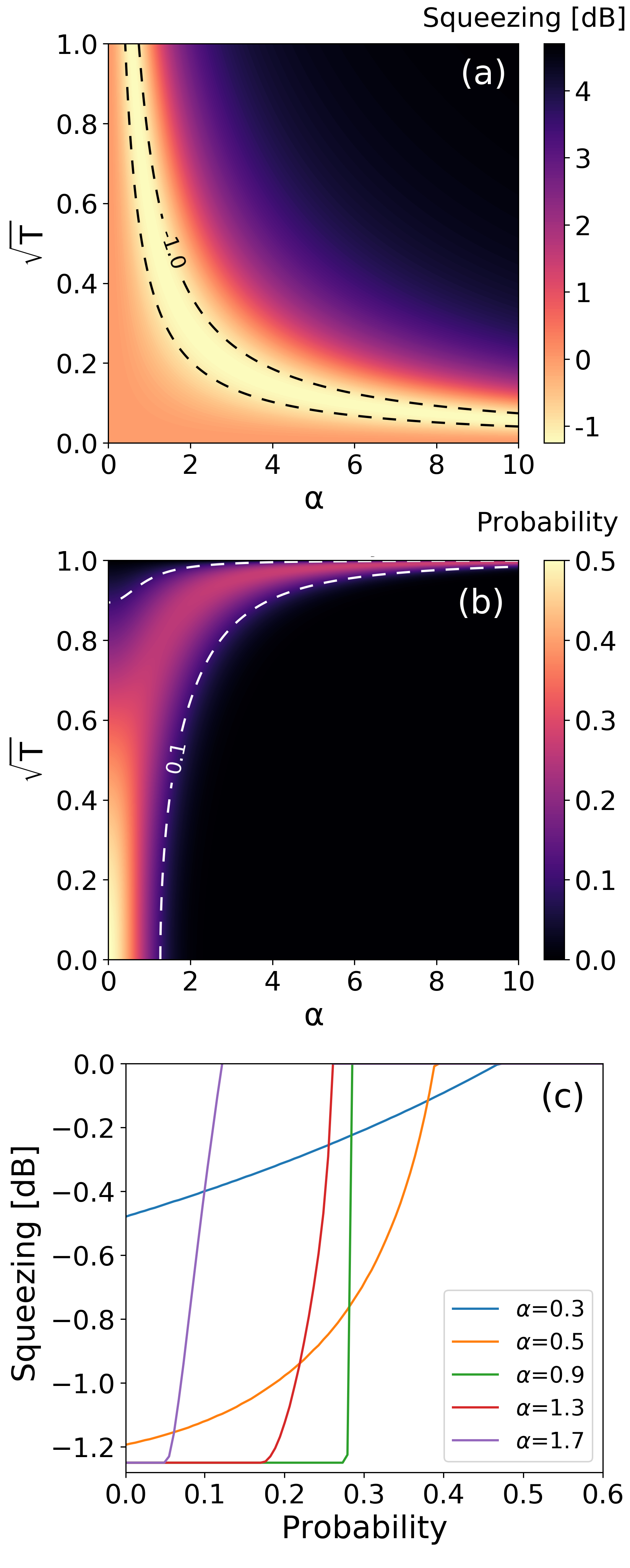}
\end{center}
\caption{(a) The squeezing and (b) the detection probability which are given by Eqs.~(\ref{eq:quadr_definition}) and (\ref{eq:simlf_det_prob}) as functions of $\alpha$ and $T$. (c) Maximally achieved squeezing for different $\alpha$ as a function of the constrained detection probability. Interferometer with fixed parameters $T_{1} = \frac{1}{2}, \ T_{2}=T_{3} \equiv T \ \textrm{and} \ T_{4} = 1$ is considered.} \label{fig:tms_vs_prob_analyt}
\end{figure}

For phases $\xi$ and $\phi$, squeezing of quadrature $C_{1}$ is maximized when $\phi = \pi/2$ and $\xi = \pm \pi$. The resulting two-mode squeezing as a function of detection beam splitter transmissivity $T$ and coherent state amplitude, along with the detection probability, are shown in Figs.~\ref{fig:tms_vs_prob_analyt}(a) and (b), respectively. 

As can be seen, the maximum squeezing is -1.25 dB and it can be achieved for all values of $\alpha$ with the appropriate choice of beamsplitter transmissivity $T$. This amount of squeezing is identical to the single-mode case in Ref.~\cite{Bartley}, and the fact that the maximum can always be achieved, independent of $\alpha$, is also similar behavior. However, for larger values of $\alpha$ the probability is strongly reduced. The maximum amount of squeezing as function of the detection probability in this case is shown in Fig.~\ref{fig:tms_vs_prob_analyt}(c). It clearly shows that the maximum squeezing that can be obtained is limited to -1.25 dB, and that the largest probability with which the maximum squeezing can be obtained gradually decreases for $\alpha>1$.


\subsection{General case}
\subsubsection{Phase dependence}
\begin{figure}[tb]
\begin{center}
\includegraphics[width=0.48\textwidth]{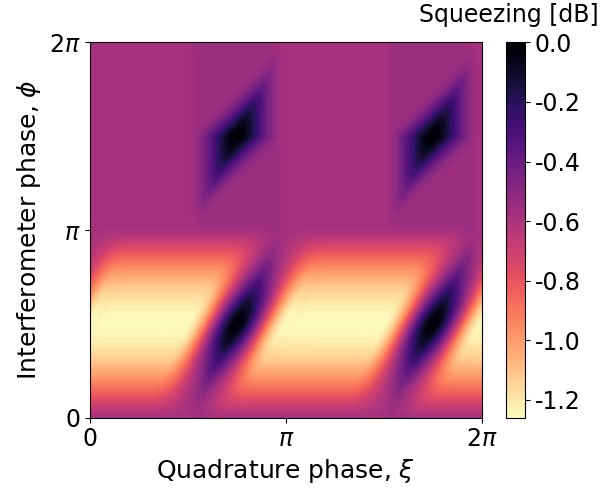}
\end{center}
\caption{The squeezing of the quadrature $C_{1}$ maximized over all beam splitter parameters $T_i$ as a function of the phases $\phi$ and $\xi$. The single PNR detection case with a critical probability $P_{\mathrm{crit}} = 0.1$ is considered and the coherent state has $\alpha=1$.} \label{fig:TMS_max_vs_phases}
\end{figure}

In general, the maximum amount of squeezing is sensitive to the interplay of phases $\phi$ and the quadrature phase $\xi$. The dependence of the squeezing on each phase, for $|\alpha|^2=1$ and maximized over all beam splitter parameters for the quadrature $C_{1}$ for a single detector registering a photon, is shown in Fig.~\ref{fig:TMS_max_vs_phases}. To perform the optimization, we use the procedure described in section~\ref{sec:optimisation} with a probability constraint of $P_{\textnormal{crit}}=0.1$. As can be seen from Fig.~\ref{fig:TMS_max_vs_phases}, to observe the maximum  squeezing, which is equal to  -1.25 dB, the phases should be set to $\phi=\pi/2, \ \xi=\pi/2 \pm \pi$. For the case where both detectors register a single photon, the maximized squeezing has a dependence very similar to that of Fig.~\ref{fig:TMS_max_vs_phases}, however, a smaller amount of squeezing, the maximum TMS is equal to -0.96 dB, can be generated. The maximum possible value of -1.25 dB squeezing is not achieved in this case, since the probability to do so does not exceed $P_\textrm{crit}=0.1$ for $|\alpha|^2=1$.

\subsubsection{Probability considerations}
\begin{figure}[tb]
\begin{center}
\includegraphics[width=0.49\textwidth]{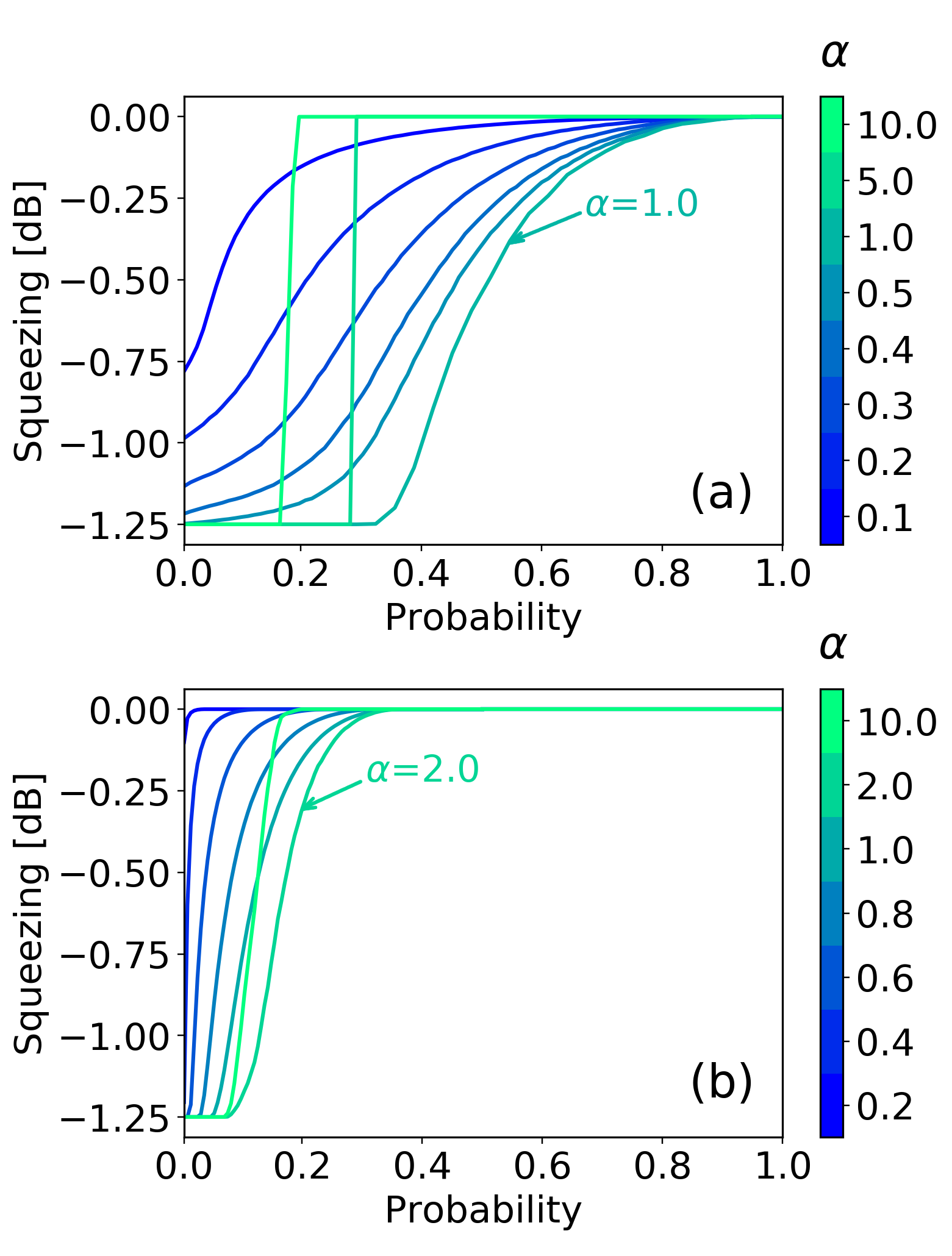}
\end{center}
\caption{The squeezing maximized over all parameters as a function of the detection probability for PNR  detection. (a) corresponds to the clicking of a single detector whereas in (b) both detectors click. The input states are a single-photon state and a coherent state with different $\alpha$ as given by the color bar.} \label{fig:tms_det_pnr}
\end{figure}

Since we explore a probabilistic effect it is important to understand which amount of squeezing can be achieved for different probability constraints. Fig.~\ref{fig:tms_det_pnr} shows the squeezing maximized over all $T_{i}$ and phases $\phi$ and $\xi$ as a function of the probability constraint for different values of $\alpha$. For a single-channel detection event, see Fig.~\ref{fig:tms_det_pnr}(a),
a value of $\alpha \approx 1$ gives the optimal result, i.e., the maximum squeezing with the largest probability.
For the case that both detectors click, see Fig.~\ref{fig:tms_det_pnr}(b), we find that some minimal $\alpha \approx 0.4$ is required to be able to achieve the maximum squeezing at all 
and that the optimal $\alpha$ is increased to about 2.
So for both cases in order to realize the maximal squeezing with a reasonable probability $\alpha$ should not be too small but should also not be very large as the maximal probability decreases with increasing $\alpha$.

\section{Click detection}\label{sec:click}

We now consider the case where click detections, rather than those with photon number resolution, are used. Analogously to our previous analysis, one can derive analytical expressions for the output states and the detection probabilities for the case of click detection in a single channel and in both channels, which are presented in Eqs.~(\ref{eq:click_case_state_single})-(\ref{eq:click_case_prob_both}). For the single-channel click detection event, the state is given by
\begin{equation}  \label{eq:click_case_state_single}
\begin{aligned}
& \rho_{\textrm{out}} = \frac{e^{-|\alpha_3|^2-|\alpha_4|^2}}{P_{\textrm{single}}} \sum_{k = 1}^{\infty}  \frac{1}{k!}|\alpha_{4}|^{2k - 2} |\psi_{k}  \rangle \langle \psi_{k}| \\
&| \psi_{k} \rangle = (k\gamma_{0} + \gamma_{1}\hat{a}_{1}^{\dag} + \gamma_{2}\hat{a}_{2}^{\dag})| \alpha_{1}, \alpha_{2} \rangle,  \\
\end{aligned}
\end{equation}
and the probability of generating this state is
\begin{equation}  \label{eq:click_case_prob_single}
\begin{aligned}
 P_{\textrm{single}} = &e^{-\left|\alpha_{\textrm{in}}\right|^2} \Bigg(\sum_{n_{2}, n_{3}, n_{4}=0}^{\infty}  \bigg|\frac{\tilde{\alpha}_{2}^{n_{2}-1}\tilde{\alpha}_{3}^{n_{3}-1}\tilde{\alpha}_{4}^{n_{4}-1}}{\sqrt{n_{2}!n_{3}!n_{4}!}} \\
& \times \left(g_{2}n_{2}\tilde{\alpha}_{3}\tilde{\alpha}_{4} + g_{3}n_{3}\tilde{\alpha}_{2}\tilde{\alpha}_{4} + g_{4}n_{4}\tilde{\alpha}_{2}\tilde{\alpha}_{3}\right)\bigg|^2 \\
& - \sum_{n_{2}, n_{4}=0}^{\infty} \bigg|\frac{\tilde{\alpha}_{2}^{n_{2}-1}\tilde{\alpha}_{4}^{n_{4}-1}}{\sqrt{n_{2}!n_{4}!}} \left(g_{2}n_{2}\tilde{\alpha}_{4} + g_{4}n_{4}\tilde{\alpha}_{2}\right) \bigg|^2 \Bigg)~, 
\end{aligned}
\end{equation}
where
\begin{equation}\nonumber
\begin{aligned}
g_{2} =& i r_{1} t_{2} \\
 g_{3} =& \gamma_{0} \\
 g_{4} =& t_{1} t_{3} \\
  \tilde{\alpha}_{2} =& \alpha_{\textrm{in}} t_{1} t_{2} \\
  \tilde{\alpha}_{3} =& \alpha_{4}  \\
  \tilde{\alpha}_{4} =& i \alpha_{\textrm{in}} r_{1} t_{3}~.  \\
\end{aligned}
\end{equation}
For click detection in both channels, the output state is given by
\begin{equation}  \label{eq:click_case_state_both}
\begin{aligned}
 \rho_{\textrm{out}} =& \frac{e^{-\left|\alpha_3\right|^2-\left|\alpha_4\right|^2}}{P_{both}}\\ 
& \times  \sum_{m=1}^{\infty}\sum_{n=1}^{\infty} \frac{\left|\alpha_{3}\right|^{2m - 2}\left|\alpha_{4}\right|^{2n - 2}}{m!n!} |\psi_{m,n}  \rangle \langle \psi_{m,n}| \\
 |\psi_{m,n} \rangle =& (m\tilde{\gamma}_{3} + n\tilde{\gamma}_{4} + \tilde{\gamma}_{1}\hat{a}_{1}^{\dag} + \tilde{\gamma}_{2}\hat{a}_{2}^{\dag})| \alpha_{1}, \alpha_{2} \rangle 
\end{aligned}
\end{equation}
where
\begin{equation}\nonumber
\begin{aligned}
  \tilde{\gamma}_{3} = &\alpha_{\textrm{in}} r_{1}^2 r_{2} r_{3}  \\
  \tilde{\gamma}_{4} = &- \alpha_{\textrm{in}} t_{1}^2 r_{2} r_{3},  \\
\end{aligned}
\end{equation}
and the probability of realizing this state is
\begin{equation}  \label{eq:click_case_prob_both}
\begin{aligned}
 P_{\textrm{both}} = &1  -  e^{-|\alpha_{\textrm{in}}|^2} \Bigg(\sum_{n_{2}, n_{3}, n_{4}=0}^{\infty}  \bigg|\frac{\tilde{\alpha}_{2}^{n_{2}-1}\tilde{\alpha}_{3}^{n_{3}-1}\tilde{\alpha}_{4}^{n_{4}-1}}{\sqrt{n_{2}!n_{3}!n_{4}!}} \\
& \times \left(g_{2}n_{2}\tilde{\alpha}_{3}\tilde{\alpha}_{4} + g_{3}n_{3}\tilde{\alpha}_{2}\tilde{\alpha}_{4} + g_{4}n_{4}\tilde{\alpha}_{2}\tilde{\alpha}_{3}\right)\bigg|^2 \\
& - \sum_{n_{1}, n_{2}, n_{4}=0}^{\infty}  \bigg|\frac{\tilde{\alpha}_{1}^{n_{1}-1}\tilde{\alpha}_{2}^{n_{2}-1}\tilde{\alpha}_{4}^{n_{4}-1}}{\sqrt{n_{1}!n_{2}!n_{4}!}} \\
& \times \left(g_{1}n_{1}\tilde{\alpha}_{2}\tilde{\alpha}_{4} + g_{2}n_{2}\tilde{\alpha}_{1}\tilde{\alpha}_{4} + g_{4}n_{4}\tilde{\alpha}_{2}\tilde{\alpha}_{3}\right)\bigg|^2 \\
& + \sum_{n_{2}, n_{4}=0}^{\infty} \bigg|\frac{\tilde{\alpha}_{2}^{n_{2}-1}\tilde{\alpha}_{4}^{n_{4}-1}}{\sqrt{n_{2}!n_{4}!}} \left(g_{2}n_{2}\tilde{\alpha}_{4} + g_{4}n_{4}\tilde{\alpha}_{2}\right) \bigg|^2 \Bigg)
\end{aligned}
\end{equation}
where
\begin{equation}\nonumber
\begin{aligned}
 g_{1} =& - r_{1} r_{2} \\
  \tilde{\alpha}_{1} =& \alpha_{3}~. 
\end{aligned}
\end{equation}

For the click detection case, the amount of squeezing can be calculated numerically using quantum states given by Eqs.~(\ref{eq:click_case_state_single})-(\ref{eq:click_case_prob_both}) in the Fock basis. In the numerical evaluations the considered states are limited to a maximal photon number of 14, which is sufficient to obtain converged results for the range of $\alpha$ values considered in this section.

\begin{figure}[tb]
\begin{center}
\includegraphics[width=0.48\textwidth]{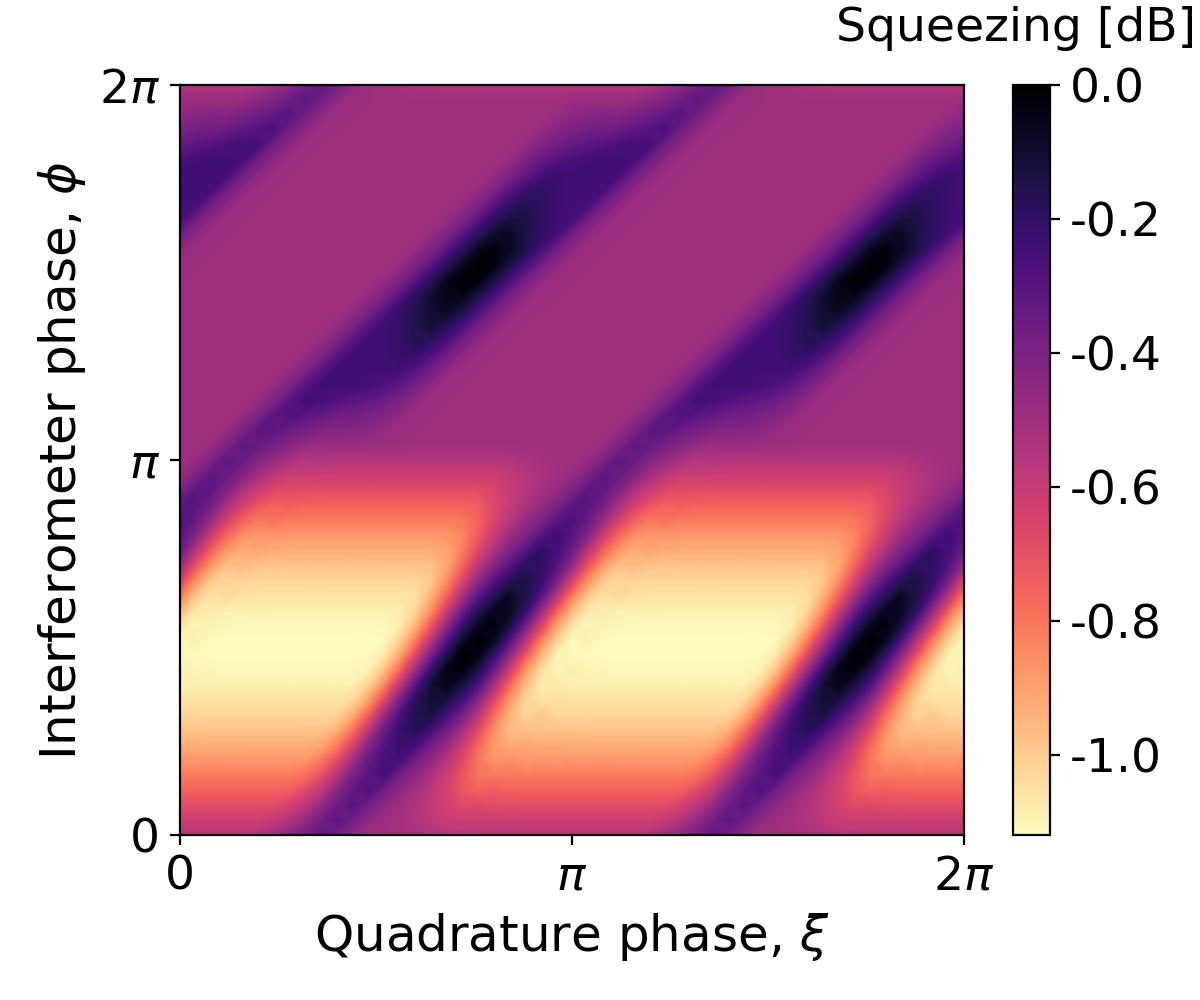}
\end{center}
\caption{The squeezed quadrature $C_{1}$ maximized over all beam splitter parameters $T_i$ as a function of the phases $\phi$ and $\xi$. The click detection case in which only one detector clicks is considered,  the critical probability is $P_{\mathrm{crit}} = 0.1$, and  the coherent state has $\alpha=1$.} \label{fig:tms_constr_vs_phases_det_click_top}
\end{figure}

\begin{figure}[tb]
\begin{center}
\includegraphics[width=0.49\textwidth]{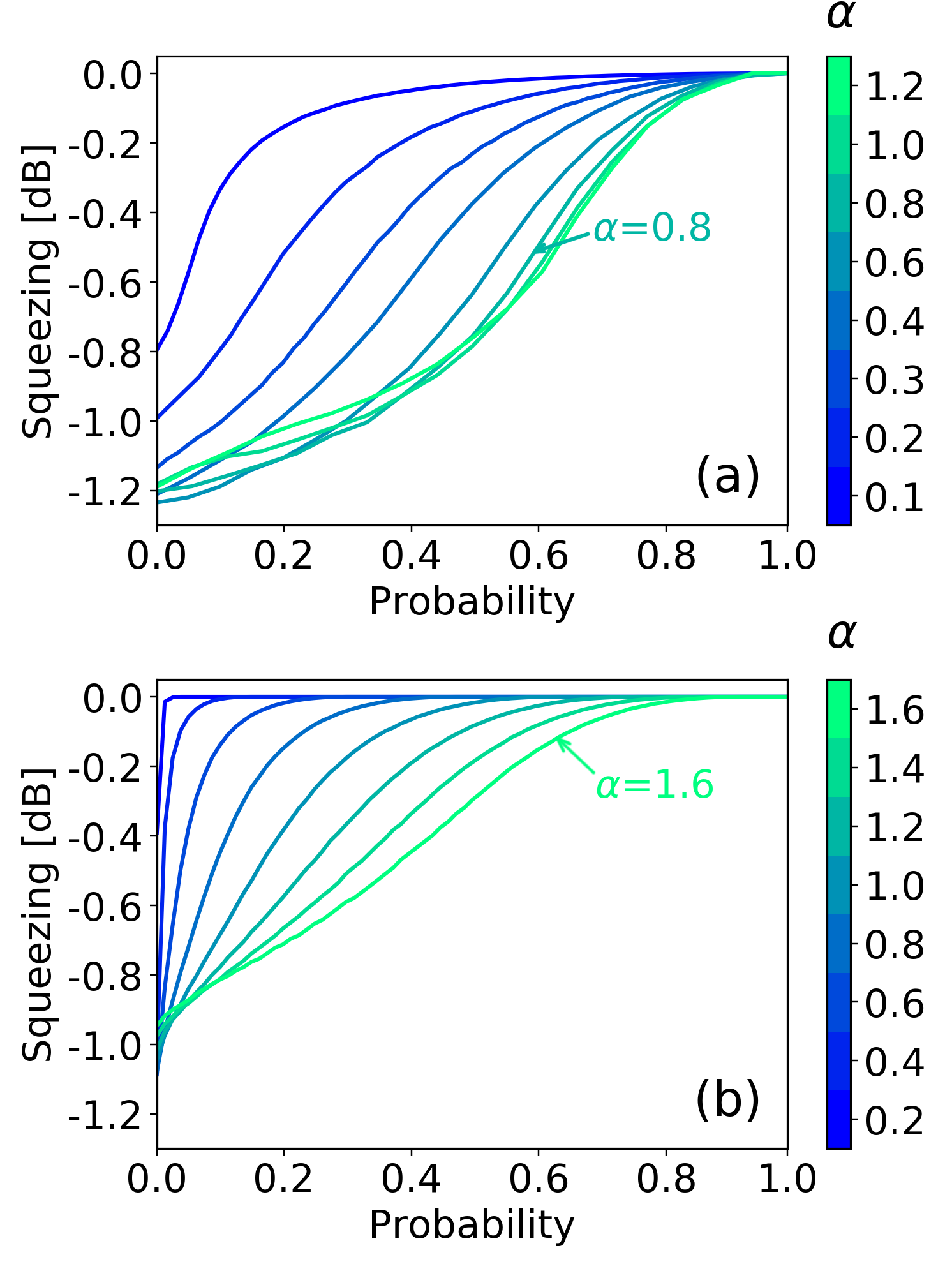}
\end{center}
\caption{The squeezed quadrature $C_{1}$ maximized over all \textbf{T} parameters as function of the detection probability for fixed phases of $\phi=\pi/2, \ \xi=\pi/2 \pm \pi$. Click detection type is considered. (a) corresponds to the clicking of a single detector whereas in (b) both detectors click. The input states are a single-photon state and a coherent state with different $\alpha$ as given by color bar. For the chosen phases the quadrature $C_{2}$ is not squeezed.} \label{fig:tms_vs_prob_det_click}
\end{figure}

\subsubsection{Phase dependence}
The maximized squeezing with the probability constraint  of $P_{\textnormal{crit}}=0.1$ as function of the phases $\phi$ and $\xi$ is shown in Fig.~\ref{fig:tms_constr_vs_phases_det_click_top}.
Compared to the PNR detection, with click detection the maximally achievable squeezing is reduced slightly
. With click detectors, the phase dependence of the maximized squeezing is quite similar to the case of PNR detection, see Fig.~\ref{fig:TMS_max_vs_phases}, and also the maximum squeezing is obtained for the phases $\phi=\pi/2, \ \xi=\pi/2 \pm \pi$ which is equal to -1.11~dB for a single detector click and -0.86~dB (not shown in figure) when both detectors click. These values are only slightly smaller than for the PNR detection case since we consider a coherent state with $\alpha=1$ for which the contributions from higher photon numbers are small.


\subsubsection{Probability considerations}

Fig.~\ref{fig:tms_vs_prob_det_click} presents  the quadrature squeezing $C_{1}$ maximized over all $T_{i}$ with the fixed phases $\phi=\pi/2, \ \xi=\pi/2 \pm \pi$ as a function of the probability constraint for different values of $\alpha$. When a single detector clicks, see Fig.~\ref{fig:tms_vs_prob_det_click}(a), the region where squeezing exceeding -1~dB can be achieved with a reasonable probability increases with increasing $\alpha$. However, already for $\alpha>0.8$ the maximum squeezing decreases with increasing $\alpha$. Thus, similar to the case of PNR detection, also for click detection a trade-off between squeezing and the detection probability is obtained.

When both detectors click, see Fig.~\ref{fig:tms_vs_prob_det_click}(b), much larger values of $\alpha$ are required to get significant squeezing with reasonable probability, since two photons are removed. In this sense, the optimal situation is reached for some $\alpha>1.6$. Although we have not computed squeezing for larger values of $\alpha$, in order to keep the numerical requirements within reasonable limits, our analysis suggests that squeezing can be obtained for all $\alpha$, however the probability is likely to be very small.

\section{Visualizing the generated states}
\begin{figure}[!thb]
\begin{center}
\includegraphics[width=0.47\textwidth]{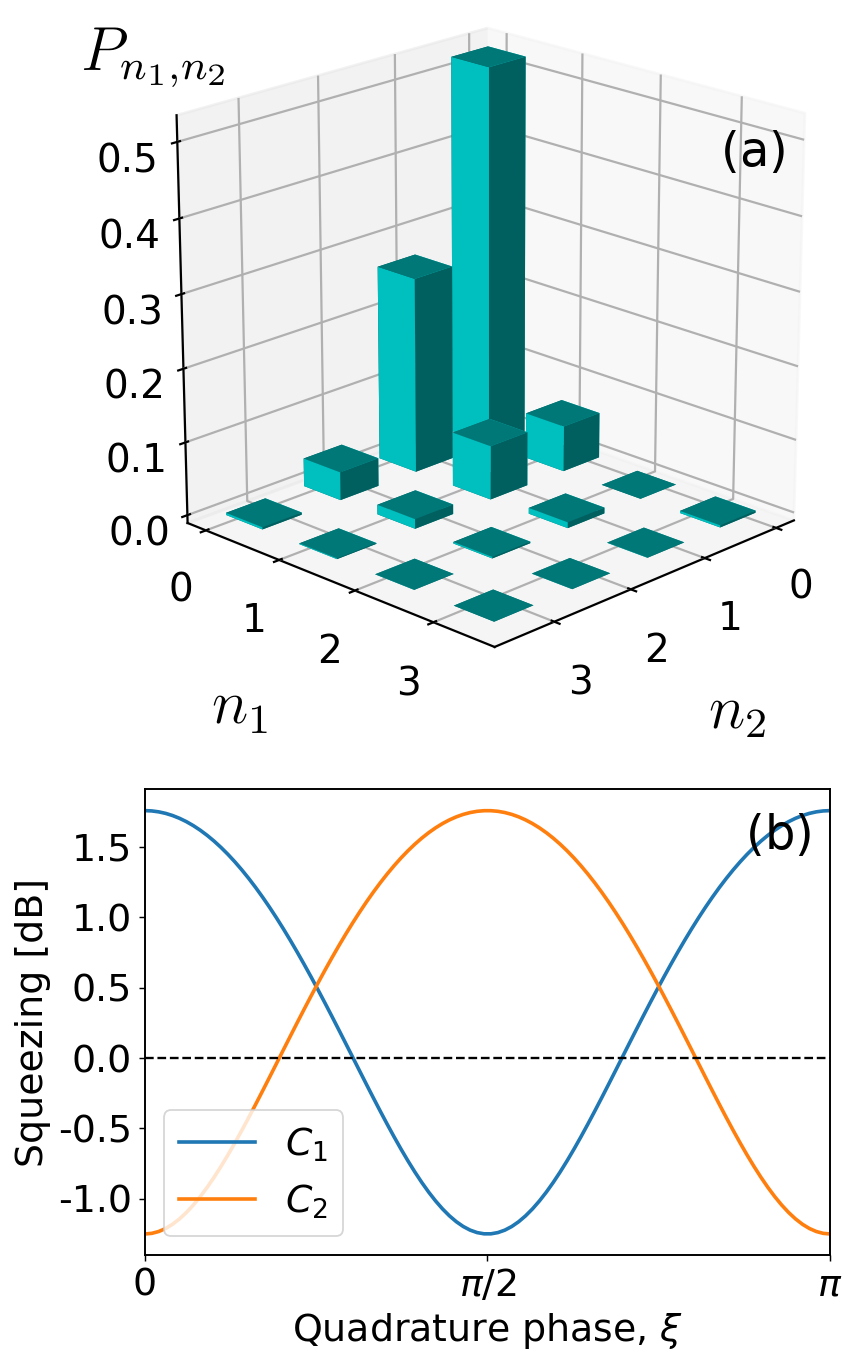}
\end{center}
\caption{Visualization of the quantum state. (a) The photon number probability distribution $P_{n_{1},n_{2}}$ between the two channels ($n_1$ and $n_2$ are the number of photons in the channels)  for maximally squeezed light with parameters: $\alpha = 1.0, \ \textbf{T} = [0.68, 0.82, 0.38, 1.0],$ $\ \phi=3\pi/2$, $\ \xi=\pi/2$. The corresponding maximum squeezing and detection probability are $S = -1.25 \ \textnormal{dB}$ and $P_{\textnormal{det}} = 0.3$, respectively. (b) Quadrature squeezing $C_1$ and $C_2$ as a function of the quadrature phase for the quantum state generated with the same parameters as in (a). Single PNR type of detection is considered.} \label{fig:phot_count_distr}
\end{figure}

In addition to quadrature squeezing, the structure of the output light can be revealed from the photon number distribution between the two channels $P_{n_{1},n_{2}} = \rho_{n_{1},n_{2},n_{1},n_{2}}$. 
As an example, the photon number distribution for maximal squeezing in the single PNR detection case with parameters $\alpha = 1.0, \ S = -1.25~\textnormal{dB}, \ \textbf{T} = [0.68, 0.82, 0.38, 1.0], \ \phi=3\pi/2,
\ \xi=\pi/2, \ P_{\textnormal{det}} = 0.3 $ is shown in Fig.~\ref{fig:phot_count_distr}(a).
Furthermore, Fig.~\ref{fig:phot_count_distr}(b) shows the dependencies of the squeezing in the two quadratures $C_1$ and $C_2$ as function of the quadrature phase. The squeezing has a sinusoidal dependence on the phase and the results for the two quadratures are phase shifted by $\pi/2$ with respect to each other, as would be expected. From this plot it is clear that the state generated is not a minimum uncertainty state, since $C_1+C_2>0$ (on a dB scale).

\begin{figure*}[!th]
\includegraphics[width=\textwidth]{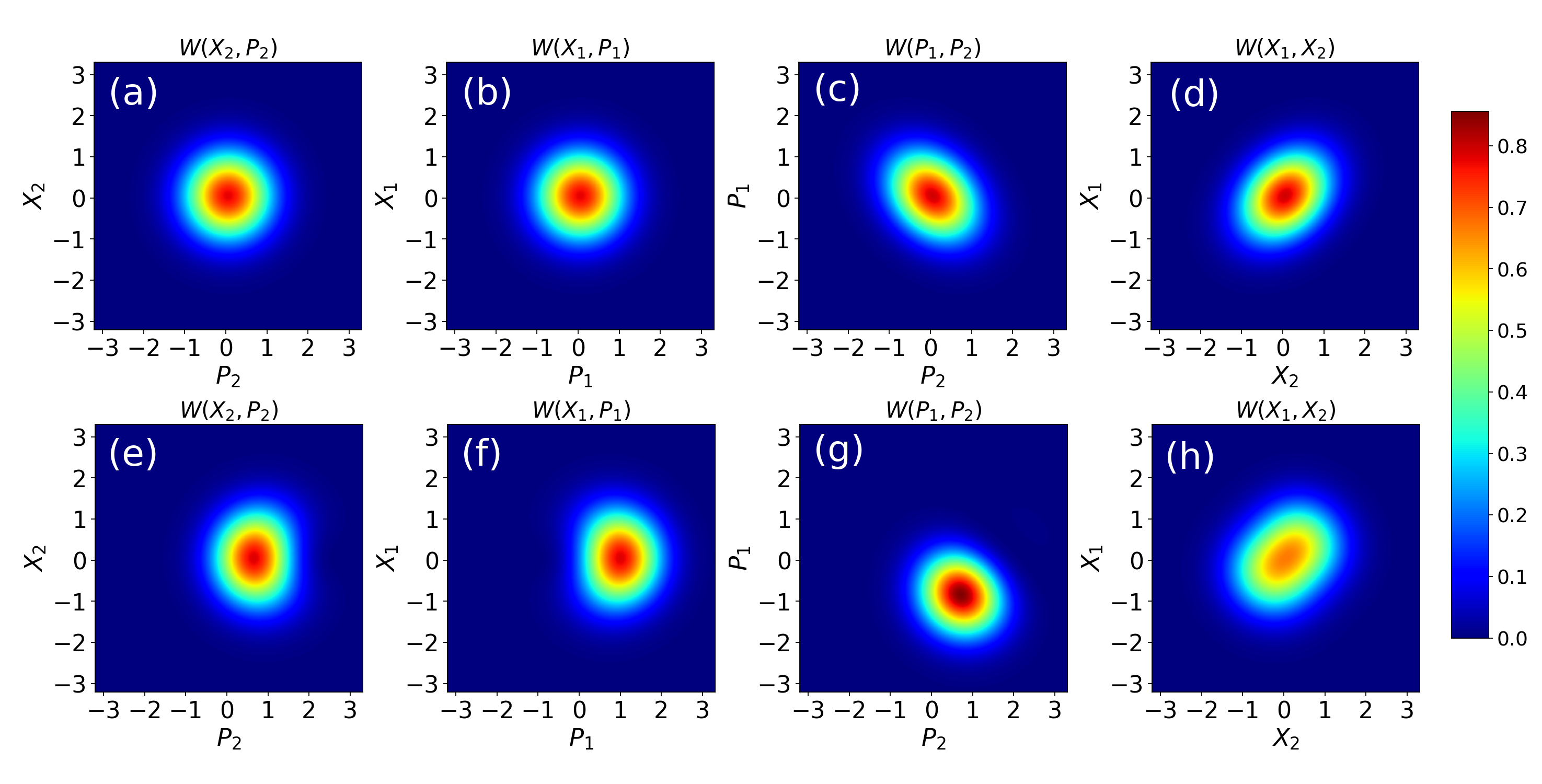}
  \caption{The reduced Wigner functions calculated using Eq.~(\ref{eq:phase_operator}). (a)-(d) correspond to the reduced Wigner functions of the TMSV state  with $z=0.143$. (e)-(h) correspond to the reduced Wigner functions of maximally squeezed light generated in the interferometer with the same parameters as in  Fig.~\ref{fig:phot_count_distr}. The corresponding squeezing and detection probability are $S_x = -1.25 \ \textnormal{dB}$ and $\textnormal{P} = 0.3$, respectively.}
  \label{fig:2d_wigner_both}
\end{figure*}

In addition to the photon number distribution, the Wigner function can give information on the phase-space distribution of the state. As an example we compare the reduced Wigner functions, see
Eq.~(\ref{eq:phase_operator}) in the Appendix, of the two-mode squeezed vacuum state (TMSV) $| \textnormal{TMSV} \rangle = \sqrt{1 - |z|^2} \sum_{n=0}^{\infty} z^n |n\rangle_{1}|n\rangle_{2} $ and the state generated in our interferometer, see Fig.~\ref{fig:2d_wigner_both}. We choose the parameter $z=0.143$ to obtain the same amount of squeezing  $S_x = -1.25 \ \textnormal{dB}$ in both cases. In the interferometer we consider a single PNR detection  with the same parameters as in Fig.~\ref{fig:phot_count_distr}. The reduced Wigner functions in Figs.~\ref{fig:2d_wigner_both}(a), (b), (e), and (f) are presented in variables of a single mode, whereas shape of the reduced Wigner functions $W(P_{1}, P_{2})$ and $W(X_{1}, X_{2})$ in Figs.~\ref{fig:2d_wigner_both}(c), (d), (g), and (h) 
is responsible for the squeezing between the two channels. When these Wigner functions take elliptic (squeezed) form, they visualize the squeezing; the more the light is squeezed the narrower the ellipses are. Moreover, one can observe that the Wigner function in Fig.~\ref{fig:2d_wigner_both}(g) is shifted relative to the origin due to the presence of a coherent state in the generated light, contrary to the TMSV in Fig.~\ref{fig:2d_wigner_both}(c).

\section{Influence of losses}
\begin{figure}[b!]
\begin{center}
\includegraphics[width=0.49\textwidth]{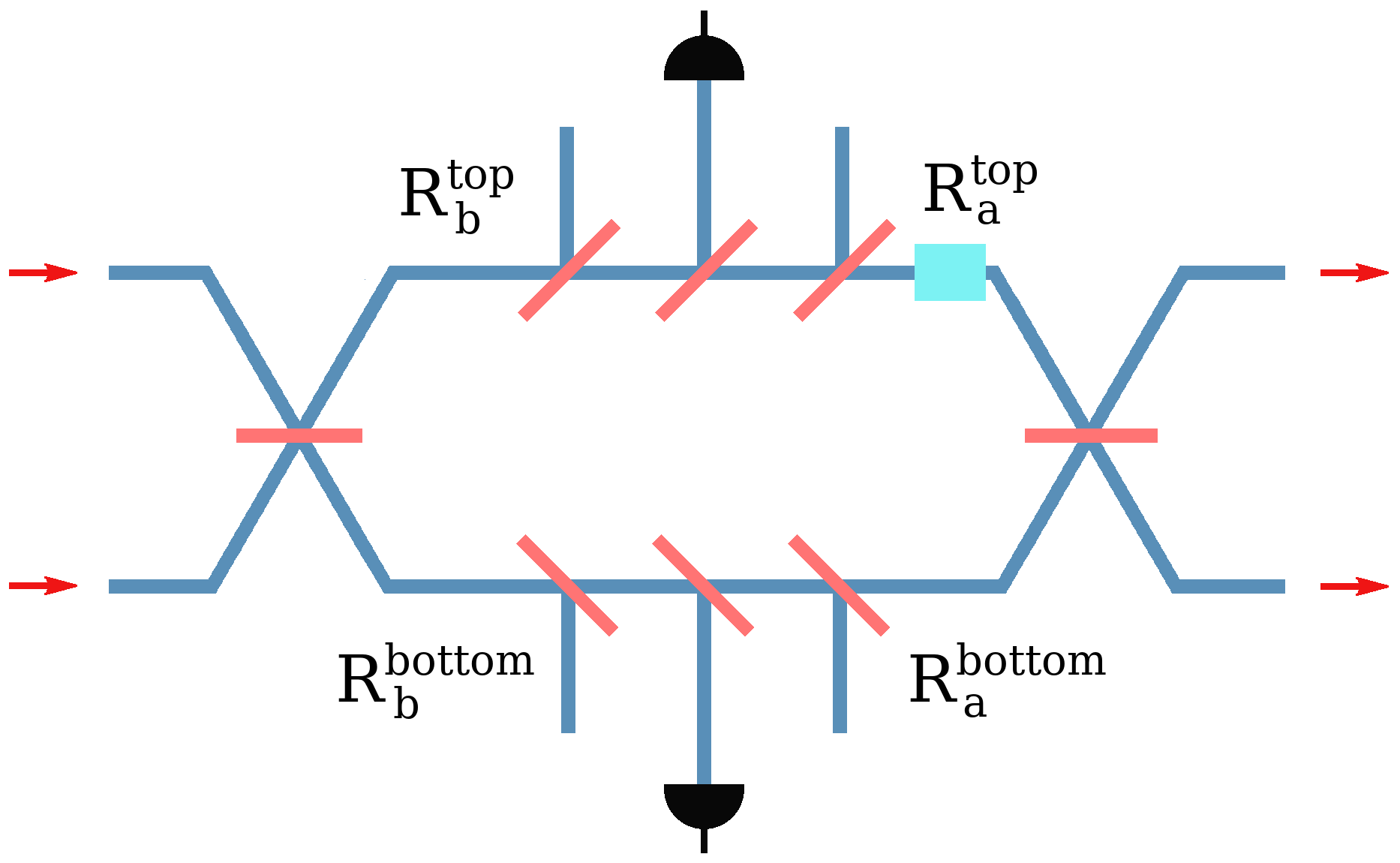}
\end{center}
\caption{A schematic representation of modeling losses by  additional beam splitters in the top and bottom channels before and after detection.\\
} \label{fig:scheme_with_losses}
\end{figure}

For the considered setup, losses related to absorption and scattering are expected to be the largest contribution. To model losses in our scheme, we place additional  beam splitters in both channels between $\mathrm{BS}_{1}$ and $\mathrm{BS}_{4}$ and consider losses before and after detection, i.e., before and after $\mathrm{BS}_{2}$ and $\mathrm{BS}_{3}$, as shown in Fig.~\ref{fig:scheme_with_losses}.

Non-vanishing reflectivities of the additional beam splitters correspond to the removal of a certain fraction of photons from our circuit. The coefficients $R_{\mathrm{b}}^{\Sigma}$ and $R_{\mathrm{a}}^{\Sigma}$ are the total reflection coefficients of the additional beam splitters (losses) placed before and after detection, respectively. They are defined as the sum of the top and bottom reflection coefficients: $R_{\mathrm{b}}^{\Sigma} = R_{\mathrm{b}}^{\mathrm{top}} + R_{\mathrm{b}}^{\mathrm{bottom}}$ and $R_{\mathrm{a}}^{\Sigma} = R_{\mathrm{a}}^{\mathrm{top}} + R_{\mathrm{a}}^{\mathrm{bottom}}$.

\begin{figure}[tb]
\begin{center}
\includegraphics[width=0.45\textwidth]{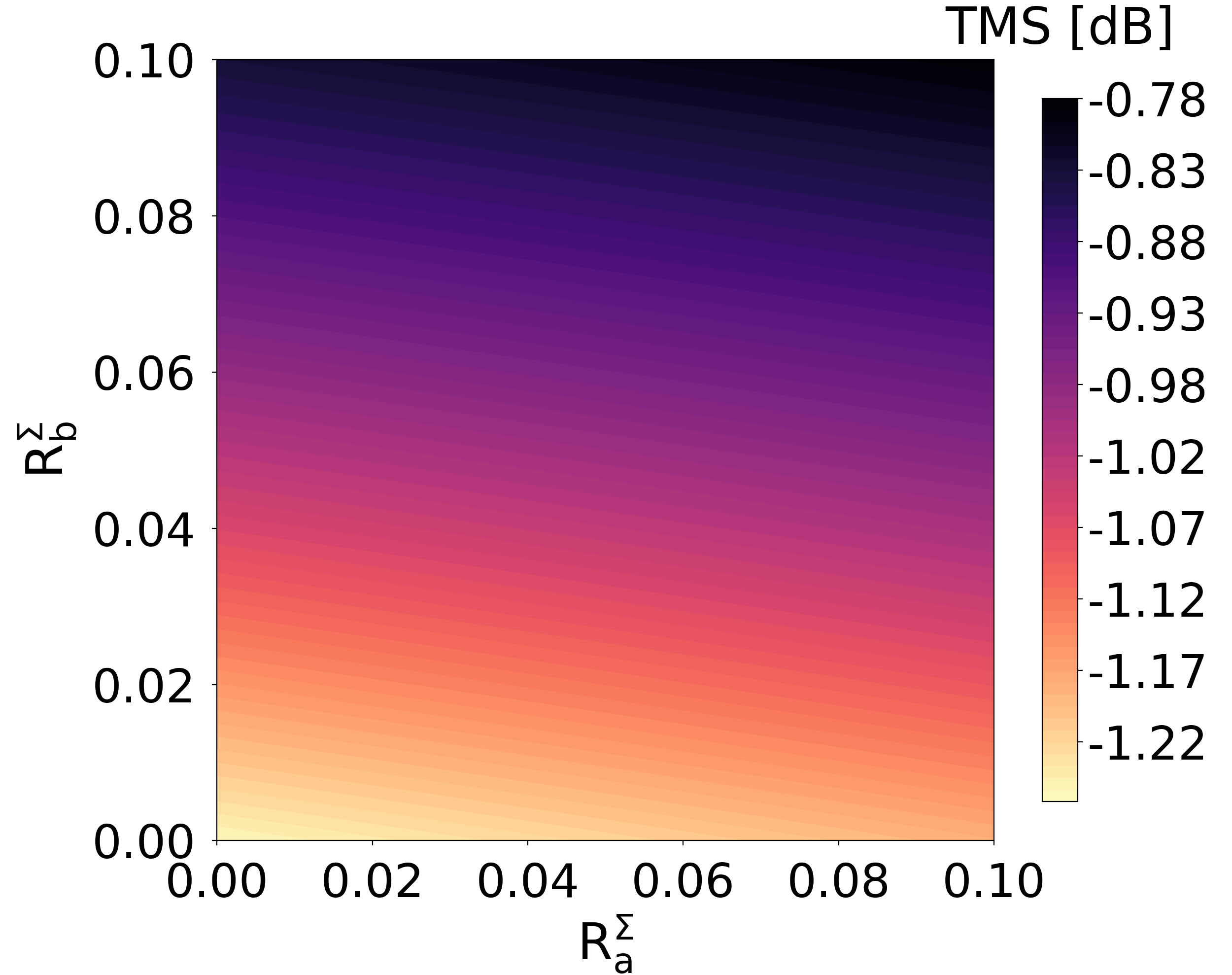}
\end{center}
\caption{Dependence of squeezing on losses in the channels. $R_{\mathrm{b}}^{\Sigma}$ and $R_{\mathrm{a}}^{\Sigma}$ are total reflection coefficients of loss BS placed before and after detection respectively defined as the sum of the top and the bottom reflection coefficients: $R_{\mathrm{b}}^{\Sigma} = R_{\mathrm{top}}^{\mathrm{before}} + R_{\mathrm{bottom}}^{\mathrm{before}}$ and $R_{\mathrm{a}}^{\Sigma} = R_{\mathrm{top}}^{\mathrm{after}} + R_{\mathrm{bottom}}^{\mathrm{after}}$, where $R_{\mathrm{top}} = R_{\mathrm{bottom}}$.} \label{fig:tms_losses}
\end{figure}

We perform numerical simulations where the coefficients $R_{\mathrm{b}}^{\Sigma}$ and $R_{\mathrm{a}}^{\Sigma}$ are varied under the condition: $R^{\mathrm{top}} = R^{\mathrm{bottom}}$ both before and after detection. For low losses, $R_{\mathrm{b/a}}^{\Sigma} \in [0, 0.1] $, the dependence of squeezing on the total reflection is shown in Fig.~\ref{fig:tms_losses}. This loss regime is compatible with state of the art implementations of integrated interferometers, in which internal circuit losses of a few percent are feasible~\cite{Sharapova}. Together with losses at the beam splitters and the detectors \cite{Ferrari} we consider 10\% as a realistic upper boundary for losses in each channel. For instance, including $5\%$ loss before and after detection ($ R_{\mathrm{b}}^{\Sigma} =R_{\mathrm{a}}^{\Sigma} = 0.05 $), squeezing is reduced from -1.25~dB to -1.0~dB. It is worth to note that losses before detection reduce the squeezing much more significantly than losses after detection, perhaps due to the different photon numbers before and after detection. For instance, $5\%$ loss before detection ($ R_{\mathrm{b}}^{\Sigma} = 0.05, R_{\mathrm{a}}^{\Sigma} = 0 $) reduces squeezing from -1.25~dB to -1.01~dB, however, including $5\%$ loss only after detection ($ R_{\mathrm{b}}^{\Sigma} = 0, R_{\mathrm{a}}^{\Sigma} = 0.05 $), squeezing is reduced from -1.25~dB to -1.21~dB.\\

\section{CONCLUSIONS}


We present theoretical and numerical investigations of a linear two-mode interferometer with nonlinear detection operations.
With a single-photon Fock state and a coherent state as the two input states to the interferometer, we analyze the influence of detection on two-mode squeezing for the cases of photon-number-resolving and click detection.
It is demonstrated that by applying detection it is possible to generate two-mode squeezing. The largest amount of squeezing that can be generated is 1.25 dB, independent of the amplitude of the coherent state, although varying this, along with other parameters of the interaction, has a significant influence on the success probability, namely that the correct measurement outcome is obtained. To investigate the feasibility of observing the predicted effects in experiments, we analyze the influence of losses and show that squeezing is degraded only weakly for not too high losses.

It is interesting to note that the amount of two-mode squeezing this interaction generates is identical to the single-mode case~\cite{Bartley}. This suggests that this interaction, when the parameters are correctly chosen, produces not only coherence between the photon-number terms required for single-mode squeezing, but also correlations between the photon-number terms when considering a two-mode state. It remains to be seen if further non-Gaussian operations acting on the modes can increase this squeezing further.

\newpage
\acknowledgments
Financial support of the Deutsche Forschungsgemeinschaft (DFG) through project number 231447078 (TRR 142, project C06) is gratefully acknowledged.
P.~R.~Sh. thanks the state of North Rhine-Westphalia for support by the
{\it Landesprogramm f{\"u}r geschlechtergerechte Hochschulen}.
We also thank the PC$^2$ (Paderborn Center for Parallel Computing) for providing computing time.

\appendix


\section{Analytical results}
\label{appa}
\subsection{No detection}
The output state of light for the circuit without detection ($T_{2}=T_{3}=1$) is given by 
\begin{equation}  \label{eq:apend_A_state}
\begin{aligned}
&| \psi \rangle =(\gamma_{01}\hat{a}_{1}^{\dag} + \gamma_{02}\hat{a}_{2}^{\dag})| \alpha_{01}, \alpha_{02} \rangle \\
&\gamma_{01} = t_{1} t_{4} - r_{1} r_{4}  \\
&\gamma_{02} = i e^{i\phi} ( t_{1} r_{4} + r_{1} t_{4} ) \\
&\alpha_{01} = i \alpha_{\textrm{in}} (t_{1} r_{4}  + r_{1} t_{4}) \\
&\alpha_{02} = \alpha_{\textrm{in}} e^{i \phi} (t_{1} t_{4}  - r_{1} r_{4}) \\
\end{aligned}
\end{equation}
where $|\alpha_{1}, \alpha_{2} \rangle = |\alpha_{1}\rangle_{1} \otimes |\alpha_{2} \rangle_{2}$ is a product of two coherent states in different channels. For this state, the variance $\Delta^{2} C_{1}=\Delta^{2} C_{2}$ can be calculated analytically and is given by
\begin{equation}  \label{eq:apend_C__fidelity}
\begin{aligned}
& \Delta^{2} C_{1} = \langle C_{1}^2 \rangle - \langle C_{1} \rangle^2 = \frac{1}{2} + \\
& + \sin{\phi} (\frac{1}{2}(t_{1}r_{1} + t_{4}r_{4}) - t_{1}^2t_{4}r_{4} - t_{4}^2t_{1}r_{1}).
\end{aligned}
\end{equation}

\subsection{Two-mode variance}

Variances of the quadratures can be calculated as
\begin{equation}  \label{eq:quadr_variances}
\begin{aligned}
& \Delta^2 C_{1} = \frac{1}{4} [\operatorname{Re}(e^{-2i\phi}(\langle a^2 \rangle + \langle b^2 \rangle + 2\langle ab \rangle) + 2\langle a b^\dag \rangle)) +  \\ 
& +\langle aa^\dag \rangle +  \langle bb^\dag \rangle - 1] - \frac{1}{2}(\operatorname{Re}(e^{-i\phi}(\langle a \rangle + \langle b \rangle)))^2 \\
& \Delta^2 C_{2} = -\frac{1}{4} [\operatorname{Re}(e^{-2i\phi}(\langle a^2 \rangle + \langle b^2 \rangle + 2\langle ab \rangle) - 2\langle a b^\dag \rangle)) -  \\ 
& -\langle aa^\dag \rangle - \langle bb^\dag \rangle + 1] - \frac{1}{2}(\operatorname{Im}(e^{-i\phi}(\langle a \rangle + \langle b \rangle)))^2 \\
\end{aligned}
\end{equation}


For the  PNR detection case the analytical formulas for average values of the operators used in Eq.~(\ref{eq:quadr_variances}) take forms: \\

\begin{widetext}
\begin{equation}  \label{eq:oper_averages}
\begin{aligned}
& \langle a \rangle = N^2 (|\gamma_{0}|^2 \alpha_{1} + \gamma_{0}^* \gamma_{1} (|\alpha_{1}|^2 + 1) + \gamma_{0}^* \gamma_{2} \alpha_{1} \alpha_{2}^* + \gamma_{1}^* \gamma_{0} \alpha_{1}^2 + |\gamma_{1}|^2 \alpha_{1} (|\alpha_{1}|^2 + 2) + \gamma_{1}^* \gamma_{2} \alpha_{1}^2 \alpha_{2}^* + \\
& + \gamma_{2}^* \gamma_{0} \alpha_{1} \alpha_{2} + \gamma_{2}^* \gamma_{1} \alpha_{2} (|\alpha_{1}|^2 + 1) + |\gamma_{2}|^2 \alpha_{1} (|\alpha_{2}|^2 + 1)) \\
& \langle b \rangle = N^2 (|\gamma_{0}|^2 \alpha_{2} + \gamma_{0}^* \gamma_{1} \alpha_{2} \alpha_{1}^* + \gamma_{0}^* \gamma_{2} (|\alpha_{2}|^2 + 1) + \gamma_{1}^* \gamma_{0} \alpha_{1} \alpha_{2} + |\gamma_{1}|^2 \alpha_{2} (|\alpha_{1}|^2 + 1) +  \\ 
& + \gamma_{1}^* \gamma_{2} \alpha_{1} (|\alpha_{2}|^2 + 1) + \gamma_{2}^* \gamma_{0} \alpha_{2}^2 + \gamma_{2}^* \gamma_{1} \alpha_{2}^2 \alpha_{1}^* + |\gamma_{2}|^2 \alpha_{2} (|\alpha_{2}|^2 + 2)) \\
& \langle a^2 \rangle = N^2 (|\gamma_{0}|^2 \alpha_{1}^2 + \gamma_{0}^* \gamma_{1} \alpha_{1} (|\alpha_{1}|^2 + 2) + \gamma_{0}^* \gamma_{2} \alpha_{1}^2 \alpha_{2}^* + \gamma_{1}^* \gamma_{0} \alpha_{1}^3 + |\gamma_{1}|^2\alpha_{1}^2(|\alpha_{1}|^2 + 3) + \\
& + \gamma_{1}^* \gamma_{2} \alpha_{1}^3 \alpha_{2}^* + \gamma_{2}^* \gamma_{0} \alpha_{2} \alpha_{1}^2 + \gamma_{2}^* \gamma_{1} \alpha_{2} \alpha_{1} (|\alpha_{1}|^2 + 2) + |\gamma_{2}|^2 \alpha_{1}^2 (|\alpha_{2}|^2 + 1)) \\
& \langle b^2 \rangle = N^2 (|\gamma_{0}|^2 \alpha_{2}^2 + \gamma_{0}^* \gamma_{1} \alpha_{2}^2 \alpha_{1}^* + \gamma_{0}^* \gamma_{2} \alpha_{2} (|\alpha_{2}|^2 + 2) + \gamma_{1}^* \gamma_{0} \alpha_{1} \alpha_{2}^2 + |\gamma_{1}|^2 (|\alpha_{1}|^2 + 1) \alpha_{2}^2 + \\
& + \gamma_{1}^* \gamma_{2} \alpha_{1} \alpha_{2} (|\alpha_{2}|^2 + 2) + \gamma_{2}^* \gamma_{0} \alpha_{2}^3 + \gamma_{2}^* \gamma_{1} \alpha_{2}^3 \alpha_{1}^* + |\gamma_{2}|^2 \alpha_{2}^2 (|\alpha_{2}|^2 + 3)) \\
& \langle a a^{\dag} \rangle = N^2 (|\gamma_{0}|^2(|\alpha_{1}|^2 + 1) + 2\Re(\gamma_{0}^*\gamma_{1}\alpha_{1}^*(|\alpha_{1}|^2 + 2)) + 2\Re(\gamma_{0}^*\gamma_{2}\alpha_{2}^*(|\alpha_{1}|^2 + 1)) + \\
& + |\gamma_{1}|^2(|\alpha_{1}|^4 + 4|\alpha_{1}|^2 + 2) + 2\Re(\gamma_{1}^*\gamma_{2}\alpha_{2}^*\alpha_{1}(|\alpha_{1}|^2 + 2)) + |\gamma_{2}|^2 (|\alpha_{1}|^2 + 1)(|\alpha_{2}|^2 + 1)) \\
& \langle  b b^{\dag}  \rangle = N^2 (|\gamma_{0}|^2(|\alpha_{2}|^2 + 1) + 2\Re(\gamma_{0}^*\gamma_{1}\alpha_{1}^*(|\alpha_{2}|^2 + 1)) + 2\Re(\gamma_{0}^*\gamma_{2}\alpha_{2}^*(|\alpha_{2}|^2 + 2)) + \\
& + |\gamma_{1}|^2(|\alpha_{1}|^2 + 1)(|\alpha_{2}|^2 + 1) + 2\Re(\gamma_{1}^*\gamma_{2}\alpha_{1}\alpha_{2}^*(|\alpha_{2}|^2 + 2)) + |\gamma_{2}|^2(|\alpha_{2}|^4 + 4|\alpha_{2}|^2 + 2)) \\
& \langle ab\rangle = N^2 (|\gamma_{0}|^2 \alpha_{1}\alpha_{2} + \gamma_{0} \gamma_{1}^* \alpha_{1}^2 \alpha_{2} + \gamma_{0} \gamma_{2}^* \alpha_{1} \alpha_{2}^2 + \gamma_{1}\gamma_{0}^*\alpha_{2}(|\alpha_{1}|^2 + 1) + |\gamma_{1}|^2 \alpha_{2} \alpha_{1} (|\alpha_{1}|^2 + 2) + \\
& + \gamma_{1} \gamma_{2}^* \alpha_{2}^2 (|\alpha_{1}|^2 + 1) + \gamma_{2}\gamma_{0}^* \alpha_{1} (|\alpha_{2}|^2 + 1) + \gamma_{2}\gamma_{1}^*\alpha_{1}^2(|\alpha_{2}|^2 + 1) + |\gamma_{2}|^2 \alpha_{1} \alpha_{2} (|\alpha_{2}|^2 + 2)) \\
& \langle a b^{\dag}\rangle = N^2 (|\gamma_{0}|^2 \alpha_{1} \alpha_{2}^* + \gamma_{0}^*\gamma_{1} \alpha_{2}^* (|\alpha_{1}|^2 + 1) + \gamma_{0}^* \gamma_{2} \alpha_{1} (\alpha_{2}^*)^2 + \gamma_{1}^*\gamma_{0}\alpha_{1}^2\alpha_{2}^* + |\gamma_{1}|^2 \alpha_{2}^* \alpha_{1} (|\alpha_{1}|^2 + 2) + \\
& + \gamma_{1}^* \gamma_{2} \alpha_{1}^2 (\alpha_{2}^*)^2 + \gamma_{2}^* \gamma_{0} \alpha_{1} (|\alpha_{2}|^2 + 1) + \gamma_{2}^* \gamma_{1} (|\alpha_{1}|^2 + 1) (|\alpha_{2}|^2 + 1) + |\gamma_{2}|^2 \alpha_{1} \alpha_{2}^* (|\alpha_{2}|^2 + 2)),  \\
\\
\end{aligned}
\end{equation}

where parameters $\gamma_{i}  (\tilde{\gamma}_{i}$), $\alpha_{i}  (\tilde{\alpha}_{i})$, and $N \equiv N_{\mathrm{single} (\mathrm{both})}$ are defined in Eqs.~(\ref{eq:state_det_pnr_single}) and (\ref{eq:state_det_pnr_both}) for cases where single-detector (both-detectors) measures one photon.

\end{widetext}

\subsection{Wigner function for two-mode state}
We use the following definition of the Wigner function for the two-mode state~\cite{Cahill, Seyfarth}:
\begin{equation} \label{eq:phase_operator}
W_{\rho}(\alpha, \beta) = 4 \textnormal{Tr}[\rho \hat{D}_{1}(2\alpha)\hat{D}_{2}(2\beta)\hat{P}_{1}\hat{P}_{2}],
\end{equation}
where $\alpha = \frac{1}{2} (X_{1} + i P_{1}) $ and $\beta = \frac{1}{2} (X_{2} + i P_{2})$ are two complex variables corresponding to modes 1 and 2, respectively, and $\hat{D}_{j}(\alpha)=\exp(\alpha \hat{a}_{j}^{\dag} - \alpha^{*}\hat{a}_{j})$ and $\hat{P}_{j}=\exp(i\pi \hat{a}_{j} \hat{a}_{j}^{\dag})$ with $j=1,2$ are the displacement and parity operators for modes 1 and 2, respectively. In order to visualize the quantum  state we integrate the four-dimensional Wigner function over two variables and define four reduced Wigner functions as:
\begin{equation} \label{eq:phase_operator}
\begin{aligned}
& W(X_{2}, P_{2}) = \int W(X_{1}, P_{1}, X_{2}, P_{2}) d X_{1} d P_{1} \\
& W(X_{1}, P_{1}) = \int W(X_{1}, P_{1}, X_{2}, P_{2}) d X_{2} d P_{2} \\
& W(P_{1}, P_{2}) = \int W(X_{1}, P_{1}, X_{2}, P_{2}) d X_{1} d X_{2} \\
& W(X_{1}, X_{2}) = \int W(X_{1}, P_{1}, X_{2}, P_{2}) d P_{1} d P_{2}, \\
\end{aligned}
\end{equation}
where each function is normalized according to $\int |W(x, y)|^2 dxdy = 1$.

\end{document}